\documentclass[12pt]{iopart}
\usepackage{graphicx}
\begin{document}

\title[Late time behavior of cosmological perturbations in a 
single brane model]{Late time behavior of cosmological perturbations in a 
single brane model}

\author{Kazuya Koyama\dag\   
\footnote[3]{Kazuya.Koyama@port.ac.uk, kazuya@utap.phys.s.u-tokyo.ac.jp}
}
\address{\dag\ Institute of Cosmology and Gravitation, Portsmouth University, 
Portsmouth, PO1 2EG, UK}

\address{\ddag\ Department of Physics, University of Tokyo, Tokyo, 113-0033, Japan}

\begin{abstract}
We present solutions for the late time evolution of cosmological
tensor and scalar perturbations in a single-Randall-Sundrum brane 
world model. Assuming that the bulk is anti-de Sitter spacetime, 
the solutions for cosmological perturbations are derived 
by summing mode functions in Poincar\'e coordinate. 
The junction conditions imposed at the moving brane are
solved numerically. The recovery of 4-dimensional 
Einstein gravity at late times is shown by solving the
5-dimensional perturbations throughout the infinite bulk. 
We also comment on several possibilities for having deviations from 
4-dimensional Einstein gravity. 
\end{abstract}



\maketitle

\section{Introduction}
Since the possibility that we are living on a brane in a higher 
dimensional spacetime was proposed, much effort has been 
devoted to investigating the cosmology of brane worlds
\cite{review}.
A model proposed by Randall and Sundrum (RS) provided a very 
simple and interesting playground in which to investigate 
brane world cosmology \cite{RS}. 
In their model, our world is realized on a four-dimensional brane 
in five-dimensional spacetime. The remarkable feature of this model
is that the size of extra dimension could be infinite. 
A homogeneous and isotropic cosmological solution
was found \cite{B1}-\cite{B3} and then the theory of cosmological 
perturbation has been extensively investigated 
\cite{Mu}-\cite{p31} (see \cite{review-p} for a review).
The study of cosmological 
perturbation is very important because it can provide a 
means for testing brane world models using forthcoming precise 
cosmological observations. However, it turns out that the 
analysis of cosmological perturbation in brane worlds is 
extremely difficult. Although a large number of papers has been 
published on this subject, there are rather few quantitative 
predications. This is because the evolution of 
perturbations on the brane is inevitably coupled to the perturbations 
in the five-dimensional bulk. Thus we need to solve very complicated 
coupled partial differential equations with complicated boundary
conditions that arise from the brane. 

Although some progress has been made in a model where 
a higher dimensional spacetime is bounded by two branes \cite{koyama}
\cite{CMB}, 
there is still no quantitative prediction of the evolution of 
perturbations in a single-brane model. 
The crucial difficulty of the single brane model is that 
the bulk spacetime extends to an infinity. A useful method for tackling 
this problem was proposed in Ref.\cite{KS1}, \cite{KS2}
(see also \cite{DDK}, \cite{KH} and \cite{tanaka}).
One point is that the background 5-dimensional spacetime
is just the anti de Sitter (AdS) spacetime (or AdS-Schwarzshild
spacetime) in the RS model. 
Thus we can easily find general solutions for the perturbations 
throughout the bulk. A difficulty arises from the existence of the brane.
We need to select a particular solution that correctly satisfies the 
boundary conditions at the brane. In this paper, we present a 
solution for this problem using a numerical method. Then the 
late time behavior of the perturbations is derived by solving 
the 5-dimensional perturbations throughout the infinite bulk. 
In this paper the bulk spacetime
is assumed to be AdS spacetime without black holes and 
we only consider the late time behavior of the perturbations. 
The generalization of the analysis will be discussed in 
the conclusion section. 

The structure of the paper is as follows. In Section 2, we 
describe the set-up of our model. In section 3, the solution 
for background spacetime is derived. In section 4, an evolution
of tensor perturbation is considered. We describe the numerical 
method used to solve the moving boundary conditions here. 
In section 5, a more complicated evolution of scalar perturbations
is derived. In section 6, we try to interpret our result using 
the gradient expansion method. Conclusions and a discussion  
of the generalization of our work are given in Section 7.

\section{The setup}
We consider the 5D action of the RS model:
\begin{eqnarray}
S &=& \frac{1}{2 \kappa^2}\int d^5 x \sqrt{-g}
\left(
{\cal R}^5 +  12 \mu^2 \right)
- \lambda \int d^4 x \sqrt{-g_{brane}} \nonumber\\
&& + \int d^4 x \sqrt{-g_{brane}} {\cal L}_{matter},
\label{0-1}
\end{eqnarray}
where ${\cal R}^5$ is the 5D Ricci scalar, $\mu$ is the 
curvature scale of the AdS spacetime and $\kappa^2=8 \pi G_5$ 
where $G_5$ is the Newton's constant in the 5D spacetime. 
The brane has tension $\lambda$ and the induced metric on the brane 
is denoted as $g_{brane}$. The tension $\lambda$ of the brane is 
taken as $\kappa^2 \lambda=6 \mu$ to ensure that the brane becomes
Minkowski spacetime if there is no matter on the brane. 
Matter is confined to the 4D brane world and is described by the
Lagrangian ${\cal L}_{matter}$. We will assume
$Z_2$ symmetry across the brane. On the brane, the metric
is given by
\begin{equation}
ds^2_{brane}= -dt^2 + a_o(t)^2 \delta_{ij} dx^i dx^j.
\end{equation}
We will decompose the perturbations into scalar, vector 
and tensor ones defined with respect to 3-space $\delta_{ij}$
and only consider tensors and scalars in this paper. 
The energy momentum tensor of the matter confined to the
brane is given by $T^{\mu}_{\nu}=T^{\mu}_{B \: \nu} 
+ \delta T^{\mu}_{\nu}$;
\begin{equation}
T^{\mu}_{B \:\nu}= diag(-\rho,p,p,p), \quad p=w \rho,
\label{1-3}
\end{equation}
\begin{equation}
\delta T^{\mu}_{\nu} = 
\left(
\begin{array}{cc}
 -\delta \rho & \delta q_{,i} \\
 a_o^{-2} \delta q_{,i}  & \delta p \: \delta_{ij}\\
\end{array}
\right).
\label{2-1}
\end{equation}
Here the matter anisotropic stress is neglected 
for simplicity. 

\section{Background spacetime}
In this section, we briefly consider the background 
spacetime. In this paper, we only consider the 
maximally symmetric bulk spacetime, that is, we assume 
that the 
bulk spacetime is given by Anti de Sitter (AdS) 
spacetime without a black hole mass. The simplest coordinate 
system for describing AdS spacetime is given by the Poicare coordinate:
\begin{equation}
ds^2= \left( \frac{1}{\mu z} \right)^2 (dz^2 
+ \eta_{\mu \nu} dx^{\mu} dx^{\nu}).
\end{equation}
Our brane is moving in this coordinate system. 
The motion of the brane is determined by imposing the 
junction condition. The brane motion is described 
by \cite{Kraus}
\begin{equation}
z=\frac{1}{\mu a_o(t)}, \tau = T(t),
\end{equation}
where
\begin{eqnarray}
H^2 &=& \left(\frac{\dot{a_o}}{a_o}\right)^2 = \frac{\kappa^2 \mu \rho}{3}
+ \frac{\kappa^4 \rho^2}{36}, \nonumber\\
\dot{T} &=& \frac{1}{a_o} \sqrt{1+\left( \frac{H}{\mu} \right)^2},
\end{eqnarray}
where $t$ is the cosmic time on the brane and a dot 
denotes the derivative with respect to $t$. The expansion of the 
universe can be understood as the motion of the brane through 
the bulk.  

Of course, the statement that our brane is moving is a coordinate 
dependent statement. It is possible to choose coordinates 
where the position of the brane is fixed. The Gaussian-Normal (GN)
coordinate is such a coordinate. The metric is given by \cite{B1}
\begin{equation}
ds^2 = dy^2 - n(y,t)^2 dt^2 +a(y,t)^2 \delta_{ij} dx^i dx^j,
\end{equation}
where
\begin{eqnarray}
a(y,t) &=& a_o(t) \left[ \cosh \mu y -\left(1+ \frac{\kappa^2 \rho}{6 \mu} \right)
\sinh \mu y \right],
\nonumber\\
n(y,t) &=& \cosh \mu y - \left(1- \frac{\kappa^2 \rho}{6 \mu}(2 +3 w) \right)
\sinh \mu y.
\end{eqnarray}
The brane is located at $y=0$. 
This coordinate is convenient for imposing the junction conditions because 
the position of the brane is fixed.  
The junction conditions in the background spacetime are given by 
\begin{eqnarray}
\left(\frac{a'}{a} \right)_b &=&- \mu \left(1+ \frac{\kappa^2 \rho}{6 \mu} \right),
\nonumber\\
\left(\frac{n'}{n} \right)_b &=& -\mu \left(1- \frac{\kappa^2 \rho}{6 \mu}
(2+3w) \right),
\end{eqnarray}
where the subscript b is used to denote bulk quantities evaluated at the brane. 
Using the Freidmann equation, the former equation can be written as 
\begin{equation}
\left(\frac{a'}{a} \right)_b = - \mu \sqrt{1+ \left(\frac{H}{\mu} \right)^2}.
\end{equation}
Note that there is a coordinate singularity in this 
coordinate $y_c$ at the finite distance from the brane 
where $a(y_c,t)=0$. This hypersurface corresponds to the 
past Cauchy horizon of the AdS spacetime. Thus, this coordinate 
is not suitable for discussing the global structure of the solution in the bulk. 
However, in order to impose the junction conditions, we need only know 
the geometry of the bulk near the brane. Thus we will solve for the bulk 
perturbations in Poincar\'e coordinate and move to the GN coordinate when 
we impose the boundary conditions. In order to do so, we need to rewrite the 
junction conditions in the GN coordinate as conditions in Poincar\'e coordinate. 
The coordinate transformation between the two coordinate systems has been 
investigated in Ref.\cite{B3}. The transformation is very complicated, but 
we only need the formula relating the derivative
with respect to the GN coordinate to the derivative
with respect to Poincar\'e coordinate on the brane.  This is
given by the following formulae;
\begin{eqnarray}
\left( \frac{\partial \tau}{\partial y} \right)_b
&=& -\frac{1}{a_o} \frac{H}{\mu}, 
\quad 
\left( \frac{\partial z}{\partial y} \right)_b =  \frac{1}{a_o} 
\sqrt{1+ \left(\frac{H}{\mu}\right)^2},  \nonumber\\
\left( \frac{\partial z}{\partial t} \right)_b
&=& -\frac{1}{a_o} \frac{H}{\mu}, 
\quad 
\left( \frac{\partial \tau}{\partial t} \right)_b = \frac{1}{a_o} 
\sqrt{1+ \left(\frac{H}{\mu}\right)^2} .
\label{coortr}
\end{eqnarray}

\section{Tensor perturbation}
Let us begin with the tensor perturbation:
\begin{equation}
ds^2=\left(\frac{1}{\mu z} \right)^2
\left( dz^2 -d\tau^2 +(\delta_{ij} +h_{ij}) dx^i dx^j \right).
\end{equation}
We expand $h_{ij}$ as 
\begin{equation}
h_{ij} = \int \frac{d^3 k}{(2 \pi)^3} 
h(z,\tau) e_{ij} e^{i k x}, 
\end{equation}
where $e_{ij}$ is the polarisation tensor. 
The evolution equation is simply given by
\begin{equation}
\frac{d^2 h}{dz^2}-\frac{3}{z} \frac{d h}{d z}
-\frac{d^2 h}{d \tau^2 }-k^2 h=0.
\end{equation}
It is easy to find a solution 
\begin{equation}
h(z,\tau) = \int dm h(m) z^2 Z_2 (mz) 
\left(e^{i \omega \tau} + c(m) e^{-i \omega \tau} \right),
\end{equation}
where 
\begin{equation}
Z_2(m z)= H_2^{(2)}(m z) + b(m) H_2^{(1)}(mz),
\end{equation}
and $\omega^2=m^2+k^2$. 
Here $H^{(1)}$ and $H^{(2)}$ are Hankel functions of the
first kind and the second kind respectively and 
$h(m), b(m)$ and $c(m)$ are arbitrary coefficients.  
This is a general solution for tensor perturbations 
in the AdS bulk. Our spacetime is bounded by the brane.
Thus we should impose a boundary condition via a junction condition. 
In the GN coordinate, the junction condition is 
simply given by 
\begin{equation}
\left[ \frac{\partial h}{\partial y} \right]_{y=0} =0.
\end{equation}
Using the formulation equations (\ref{coortr}), we can rewrite this condition 
in terms of the Poincar\'e coordinate
\begin{equation}
\left[
\frac{\partial h}{\partial z}-\frac{H}{\mu} 
\left(1 + \left(\frac{H}{\mu}\right)^2 \right)^{-1/2}
\frac{\partial h}{\partial \tau} 
\right]_{z=1/\mu a_o(t),\tau=T(t)} =0.
\label{boun1}
\end{equation}
Let us impose the boundary condition Eq.(\ref{boun1}) 
on a general solution given by
\begin{equation}
h(z,\tau) = \int dm h(m)  m^2 z^2 Z_2 (mz) e^{i \omega \tau}.
\label{gen}
\end{equation}
The boundary condition of equation (\ref{boun1}) gives
\begin{eqnarray}
\int dm h(m) \left[  
\frac{m^3}{\mu^2 a_o^2} Z_1 \left(\frac{m}{\mu a_o} \right) 
-\frac{H}{\mu} \left( 1 + \left(\frac{H}{\mu}\right)^2  
\right)^{-1/2} \frac{ i \omega m^2}{\mu^2 a_o^2} 
Z_2 \left(\frac{m}{\mu a_o} \right)  \right] e^{i \omega T} \nonumber\\
=0. 
\label{boun}
\end{eqnarray}
The problem is finding the solution for $h(m)$ that satisfies
this boundary condition. 
In order to find the solution, we need a numerical method. 

As suggested by recent numerical works \cite{p26}, \cite{p29},
we expect the behavior of the perturbation to be well described 
by a standard 4D Einstein gravity at low energies $H/\mu \ll 1$. 
We can argue on the recovery of the 4D solution as follows. 
At late times, it is natural to assume that the 
Kaluza Klein (KK) mass is small compared with the bulk curvature 
scale because the Hubble scale of the brane universe is much 
smaller than the bulk curvature scale $H/\mu \ll 1$.  So we assume
\begin{equation}
\frac{m}{\mu a_o} \equiv \epsilon \ll 1.
\label{KKmass}
\end{equation}
Then expanding in terms of $\epsilon$ by using the asymptotic behavior
of Hankel functions with small arguments,
\begin{equation}
H^{(1)}_{\nu}(z) \sim - i \frac{(\nu-1)!}{\pi} 
\left(\frac{2}{z}\right)^{\nu},\quad 
H^{(2)}_{\nu}(z) \sim i \frac{(\nu-1)!}{\pi} 
\left(\frac{2}{z}\right)^{\nu},\quad 
\end{equation}
the general solution of equation (\ref{gen}) on the brane becomes
\begin{equation}
h_b(t) \sim \int dm h(m) e^{i \omega \eta},
\end{equation}
where we neglected the numerical factor and used $T(t)=\eta$ at 
late times where $\eta$ is the conformal time. 
On the other hand, the boundary condition of equation (\ref{boun}) becomes 
\begin{eqnarray}
0 = \int dm \left( \frac{m^2}{a_o} - 2 i \omega H 
  +{\cal O}\left(\epsilon^4, \epsilon^4 \log \epsilon  \right) 
  \right) h(m) e^{i \omega \eta} .
\label{log1}
\end{eqnarray}
Then on rewriting $m^2=k^2+\omega^2$ and $\omega$ in terms of the derivative 
with respect to conformal time, this condition can be written as 
\begin{equation}
\left( -\frac{d^2}{ d \eta^2} - 2 \frac{1}{a_o} \frac{d a_o}{d \eta} 
\frac{d}{d \eta} -k^2 \right) h_b=0.
\label{4Dt}
\end{equation}
This is noting but the evolution equation obtained in 
4D Einstein gravity for the tensor perturbation. It implies that, 
at late times $H/\mu \ll 1$,
the boundary condition selects a particular solution that
obeys the 4D evolution equation on the brane. 
It should be noted that the excitation of KK modes in 
Poincar\'e modes is necessary in order to satisfy the boundary 
condition because zero-mode solution is just $e^{i k \eta}$. 
The movement of the brane excites KK modes. They give the damping 
of zero-mode and give the friction term in Eq.(\ref{4Dt}). 

However, the above argument assumes that the junction 
condition (\ref{boun}) can be satisfied by the KK modes with the 
condition (\ref{KKmass}). It only ensures that once  
there is a solution for $h(m)$ that satisfies the condition 
(\ref{KKmass}), the 4D evolution equation is recovered.
We should check that there exists a solution for the 
junction condition (\ref{boun}) that satisfies the condition 
(\ref{KKmass}). For this purpose, the spectrum of the KK mass 
should be determined by imposing the junction  
condition at the brane (\ref{boun}) as well as the initial 
conditions and the boundary condition in the bulk.

Let us try to find solutions for $b(m), c(m)$ and $h(m)$
numerically. We first specify the 
boundary condition at the AdS past Cauchy horizon, which corresponds
to the infinity $z \to \infty$. 
The most natural assumption is that the wave is outgoing, so
that there is no incoming radiation. We demand 
\begin{equation}
h \propto \int dm z^{3/2} e^{i m (z-\tau)},
\end{equation}
for $z \to \infty$ and $k \to 0$. This condition gives $b(m)=c(m)=0$.
Then we impose the boundary condition of equation (\ref{boun}) at the brane. 
This condition is formally written as
\begin{equation}
\int dm h(m) {\cal B}(m,t) =0.
\end{equation} 
We numerically find solutions for $h(m)$ that satisfy this 
boundary condition approximately.
We first discretize time $t$ and a KK mass $m$.
Then we end up a matrix equation. 
A problem is finding eigenvectors with a null eigenvalue.
In principle there could be an infinite number of solutions
for $h(m)$. However we expect that an excited KK mass is small 
at low energies $H/\mu \ll 1$ as long as we consider an initial 
condition that does not contain a significant excitation of KK modes.  
Thus we introduce a cut-off for $m$. Then we can 
find finite numbers of approximate solutions for 
$h(m)$. In addition, we should determine the initial condition. 
Here we adopt an ad hoc initial condition that the 
perturbation is constant with respect to $t$ on the 
brane at super-horizon scales. This can be achieved by mixing the real 
part and the imaginary part of $h$ appropriately. Then we 
obtain a real solution for $h$. In appendix C, the 
accuracy of the numerical calculation is shown.

Figure 1 shows two solutions for $h(m)$ from numerical results. 
We take $H/\mu =0.0001$ at the horizon crossing as an example. 
As expected we need the KK 
modes to satisfy the boundary condition but the excited KK mass
is small compared with $\mu$. If we sum the mode functions 
with the weight $h(m)$, we recover the 4D behavior of 
the perturbation on the brane (Figure 2). 

We would like to make a comment on the cut-off of the KK modes 
introduced in the numerical calculations. As is seen from the 
Figure 1, the solution obtained for $h(m)$ is localized 
around $m/\mu < 0.2$ which is well below the artificial 
cut-off at $m/\mu=1.2$. We have confirmed the existence of the 
solution for $h(m)$ that is localized around $m/\mu < 0.2$
even if we increase the cut-off up to $m/\mu =2.4$.  
Thus the existence of the solution with $m/\mu a_o <<1$ does not 
depend on the artificial cut-off of the KK mass. 
On the other hand, the introduction of the cut-off certainly 
kills the solutions that have large $m/\mu a_o$. 
Thus our analysis is limited to the solutions with the KK mass 
below the cut-off. 

\begin{figure}[h]
\begin{center}
\includegraphics[width=15cm]{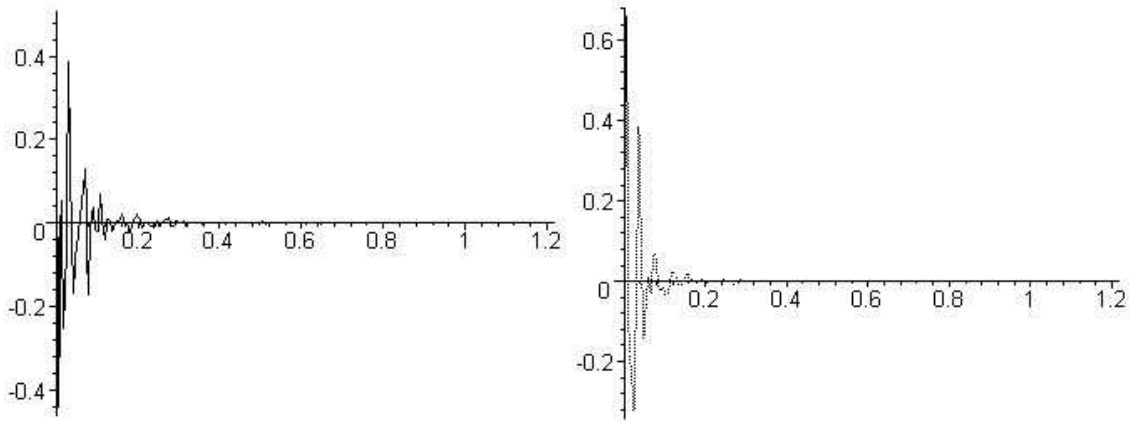}
\end{center}
\caption{Solutions for $h(m)$. The vertical line 
is $m/\mu$. We take $H/\mu=0.0001$ at the horizon crossing 
and the universe is assumed to be radiation dominated.
The cut-off is introduced at $m/\mu =1.2$.} 
\end{figure}

The different solutions for $h(m)$ correspond to different 
initial profiles in the bulk (Figure 3). In this sense, 
we have not solved the problem yet. In order to find the solution 
that corresponds to a given initial profile, we again sum solutions
with different $h(m)$. Fortunately, we find that the different 
solutions $h(m)$ give the same solution on the brane. This indicates 
that the recovery of the 4D behavior does not depend on the choice of 
the initial profile, as long as the initial profile does not 
contain large KK masses. 

\begin{figure}[h]
\begin{center}
\includegraphics[width=10cm]{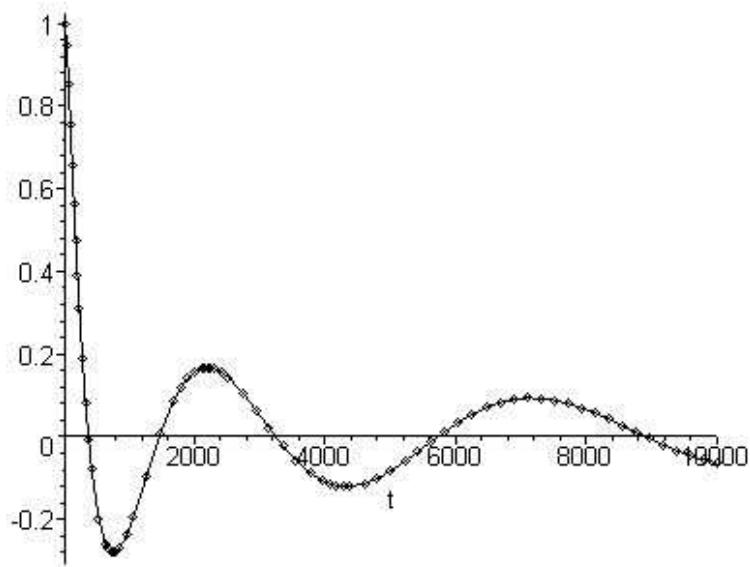}
\end{center}
\caption{Behavior of the solution $h_b$ on the brane. The horizontal axis
is the cosmic time.
The solid line is the prediction of 4D cosmology with the same
initial condition. The points represent the solutions in the brane 
world obtained by summing the Poincar\'e mode functions 
with the weight $h(m)$. 
All solution for $h(m)$ found in numerical calculations 
reproduces the 4D behavior of the perturbation.}
\end{figure}
\begin{figure}[h]
\begin{center}
\includegraphics[width=8cm]{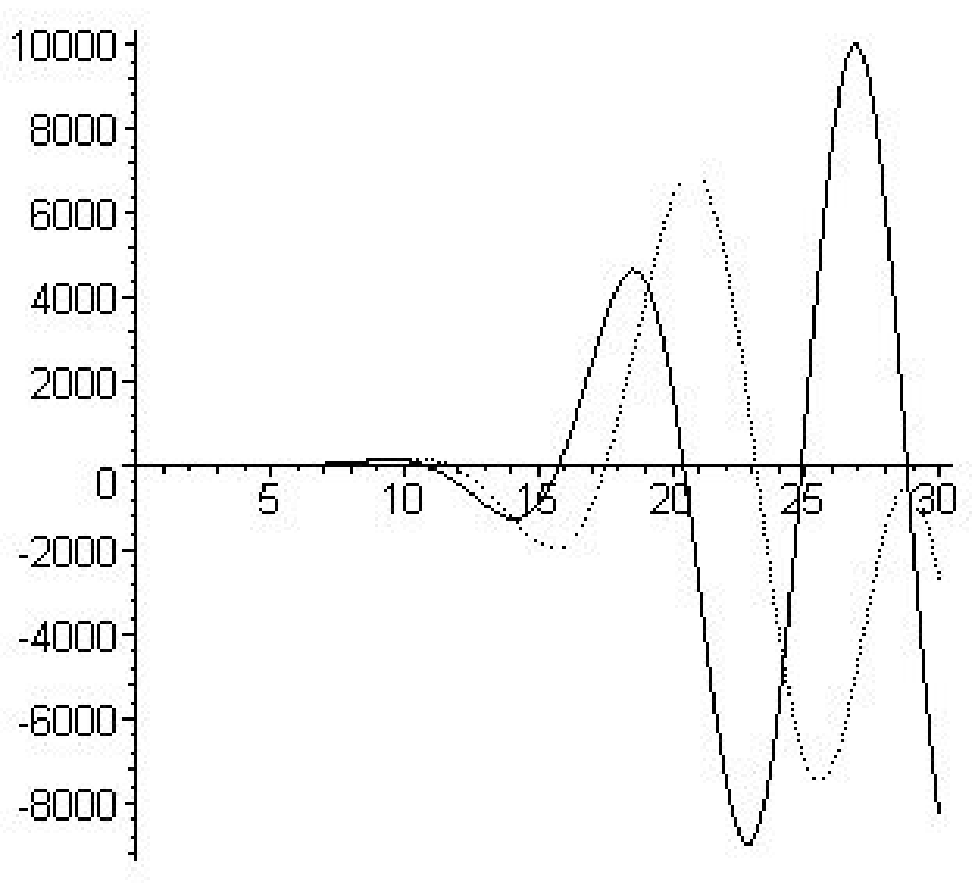}
\end{center}
\caption{The initial profile of the solution $h(t_i,z)$ in the bulk.
The horizontal axis is $\mu z$.  
The solid line corresponds to the solution derived from $h(m)$ in 
the left panel of Figure 1 and 
the dotted line corresponds to that in the right panel.}
\end{figure}

If we include the high energy corrections ${\cal O}(H/\mu)$, 
the behavior of the perturbation significantly deviates 
from 4D one. 
The difficulty of the high energy corrections is that the 
corrections depend on the choice of the initial conditions. 
Unlike the late time evolution, the different solutions for $h(m)$
give different evolutions on the brane.  Thus, unless we perform 
the summation of the solutions with different $h(m)$ we cannot get 
the final answer. As we mentioned, the different solutions for $h(m)$
correspond to different initial profiles. Thus this indicates the 
sensitivity of the solution to the choice of initial conditions.   
A detailed study of the high energy correction is beyond the scope
of this paper and it will be reported in a separate publication \cite{HKT}.

\section{Scalar perturbation}
Now let us consider the scalar perturbation. There are
several ways to calculate the scalar perturbation 
in AdS spacetime. Here we describe an approach based on 
a master variable. Using the generalized 5D longitudinal gauge, 
the perturbed spacetime is given by \cite{BDBL}
\begin{eqnarray}
ds^2 &=& b(y,t)^2(1+2 A_{yy})dy^2+n(y,t)A_{y}dydt -n(y,t)^2(1+2 A^2) dt^2 
\nonumber\\
&& +a(y,t)^2(1+2 {\cal R})\delta_{ij} dx^i dx^j. 
\end{eqnarray}
It was shown that the solution for metric perturbations can be 
derived from a master variable $\Omega$ \cite{Mu} \cite{KIS} 
\cite{p16} \cite{p21};
\begin{eqnarray}
 \label{lA}
 A &=& -\frac{1}{6a}
 \left\{ \frac{1}{b^2} \left[
  2\Omega'' - \left( 2\frac{b'}{b}+\frac{n'}{n}\right) \Omega'
  \right]
 + \frac{1}{n^2} \left[ \ddot\Omega - \left( 2\frac{\dot{b}}{b}+\frac{\dot
n}{n} \right) \dot\Omega \right]
  - \mu^2 \Omega
 \right\}
\, , \nonumber\\
\label{lAy}
A_y &=&
\frac{1}{na}\left(
 \dot\Omega'-\frac{n'}{n}\dot\Omega - \frac{\dot{b}}{b}\Omega'
  \right)\,, \nonumber\\
\label{lAyy}
 A_{yy} &=& \frac{1}{6a} \left\{
  \frac{1}{b^2} \left[
   \Omega''- \left( 2\frac{n'}{n} + \frac{b'}{b} \right) \Omega' \right]
 + \frac{1}{n^2} \left[
  2\ddot\Omega-\left( 2\frac{\dot{n}}{n} + \frac{\dot{b}}{b} \right)
  \dot\Omega\right]
 +\mu^2 \Omega \right\} \,, \nonumber\\
\label{lR}
{\cal R} &=&
 \frac{1}{6a}\left\{
 \frac{1}{b^2} \left[
  \Omega''+ \left( \frac{n'}{n}-\frac{b'}{b}\right) \Omega' \right]
 +\frac{1}{n^2}\left[
  -\ddot\Omega + \left( \frac{\dot n}{n}-\frac{\dot{b}}{b}\right) \dot\Omega
  \right]
 - 2\mu^2 \Omega\right\} \,,
 \label{metricomega}
\end{eqnarray}
as long as the master variable $\Omega$ in the bulk satisfies a 
wave equation given by
\begin{equation}
 - \left( {b\over na^3} \dot\Omega \right)^\cdot
 + \left( {n\over b a^3} \Omega' \right)^\prime
 + \left( \mu^2 + {\nabla^2\over a^2} \right)
 {nb\over a^3} \Omega
 = 0
 \,.
\end{equation}
In Poincar\'e coordinate, the master equation is given by
\begin{equation}
\frac{d^2 \Omega}{d z^2}+ \frac{3}{z} \frac{d \Omega}{d z} 
+ \frac{\mu^2}{z^2} \Omega -
\frac{d^2 \Omega}{d \tau^2}+\nabla^2 \Omega=0
\end{equation}
The solution is easily found as \cite{KIS}
\begin{equation}
\Omega = -\frac{2}{z} \mu^{-3} \int \frac{d^3 k}{(2 \pi)^3}
\int dm S(m) Z_0 (mz) e^{-i \omega \tau}
e^{i k x}.
\label{omega}
\end{equation}
A factor $-2$ was added just for later convenience. 

The junction conditions are much more complicated. 
In order to impose the junction conditions, we move
to the GN coordinate. The junction conditions have been found already 
in literatures (see \cite{KIS} and \cite{p21}), 
but, for completeness, we briefly discuss the way to 
impose junction conditions on a master variable in Appendix A. 
The junction conditions give the expressions for matter perturbations 
on the brane in terms of a master variable;
\begin{eqnarray}
a_o \kappa^2 \delta \rho &=& 
-\frac{k^2}{a^2} \left(\Omega'-\frac{a'}{a} \Omega \right)
- 3 \frac{\dot{a}}{a} \left( \dot{\Omega}'- \frac{n'}{n} 
\dot{\Omega} \right),  \nonumber\\
a_o \kappa^2 \delta q &=& 
- \left(\dot{\Omega}' - \frac{n'}{n} \dot{\Omega}  \right),
\nonumber\\
a_o \kappa^2 \delta p &=&
\ddot{\Omega}'-\frac{a'}{a} \ddot{\Omega} 
+2 \frac{\dot{a}}{a} 
\left( \dot{\Omega}'-\frac{n'}{n} \dot{\Omega} \right) 
+ \left\{4 \frac{\dot{a}}{a} \left(\frac{a'}{a}-\frac{n'}{n}
\right) + 2 \left(\frac{\dot{a}}{a}\right)' - 
\left(\frac{\dot{n}}{{n}}\right)'  \right\} \dot{\Omega}
 \nonumber\\
&-& \frac{2}{3} \left( \frac{a'}{a}-\frac{n'}{n} \right) 
\frac{k^2}{a^2} \Omega +\mu^2
\left(\frac{a'}{a}-\frac{n'}{n} \right) \Omega
-\left(\frac{a'}{a}-\frac{n'}{n}\right)
\left(2 \frac{a'}{a}- \frac{n'}{n} \right) \Omega', \nonumber\\
\label{bomega}
\end{eqnarray}
where the prime denotes $\partial_y$ and dot denotes $\partial_t$ in 
the GN coordinate and the right hand side of the equation should 
be evaluated at the brane. 
Imposing the equation of state $\delta p = c_s^2 \delta \rho$,
we get the boundary condition for $\Omega$. 

It is more convenient to rewrite these equations into 
the follwoing form using the expressions for metric perturbations
in terms of a master variable equations (\ref{metricomega}) as
\begin{eqnarray}
-\frac{1}{2} \kappa^2 {\cal H} \delta \rho &=& 
3 H (\dot{{\cal R}} -H A) + \frac{k^2}{a_o^2} {\cal R} -\frac{1}{2}
\kappa_4^2 \delta \rho_{{\cal E}}, \nonumber\\
-\frac{1}{2} \kappa^2 {\cal H} \delta p &=& 
-\ddot{{\cal R}} - H \left(3 -\frac{\dot{H}}{{\cal H}^2} \right) \dot{{\cal R}}
+H \dot{A}+\left(2 \dot{H}+ 3 H^2 - \frac{H^2 \dot{H}}{{\cal H}^2} \right) A
\nonumber\\ 
&& -\frac{1}{3} \frac{k^2}{a_o^2} A 
-\frac{1}{3} \left(1 - \frac{\dot{H}}{{\cal H}^2} \right) \frac{k^2}{a_o^2}{\cal R}
-\frac{1}{6}\left(1+\frac{\dot{H}}{{\cal H}^2}  \right) \kappa_4
\delta \rho_{{\cal E}}, \nonumber\\
-\frac{1}{2} \kappa^2 {\cal H} \delta q &=& 
\dot{{\cal R}} - H A -\frac{1}{2} \kappa_4^2 \delta q_{{\cal E}}, \nonumber\\
0 &=& \frac{1}{a_o^2} (A + {\cal R}) + \kappa_4^2 \delta \pi_{{\cal E}},
\end{eqnarray}
where we defined 
\begin{equation}
{\cal H} \equiv \left(\frac{a'}{a} \right)_b =-\mu 
\sqrt{1+\left(\frac{H}{\mu}\right)^2},\quad \kappa_4^2=\kappa^2 \mu,
\end{equation}
and 
\begin{eqnarray}
\kappa_4^2 \delta \rho_{{\cal E}} &=& 
\frac{k^4 \Omega}{3 a_o^5}, \nonumber\\
\kappa_4^2 \delta q_{{\cal E}} &=& 
\frac{k^2}{3 a_o^3} \left(\frac{\dot{a_o}}{a_o} \Omega - \dot{\Omega} \right), 
\nonumber\\
\kappa_4^2 \delta \pi_{{\cal E}} &=& 
\frac{1}{6 a_o^3} \left(3 \ddot{\Omega} - 3 H \dot{\Omega}
+\frac{k^2}{a_o^2} \Omega -3 \left(\frac{n'}{n}-\frac{a'}{a} \right)_b
\Omega' \right).
\end{eqnarray}
These equations can be derived from the projected Einstein equation on the brane
\cite{SMS};
\begin{equation}
G_{\mu \nu} + {\cal E}_{\mu \nu}
=\kappa_4^2 T_{\mu \nu} + \kappa^4 \Pi_{\mu \nu},
\label{1-1-1}
\end{equation}
where
\begin{equation}
\Pi_{\mu \nu}=-\frac{1}{4}T_{\mu \alpha}T_{\nu}^{\alpha}
+\frac{1}{12} T^{\alpha}_{\alpha}T_{\mu \nu}+
\frac{1}{24}(3 T_{\alpha \beta}T^{\alpha \beta}- 
(T^{\alpha}_{\alpha})^2)
g_{\mu \nu},
\end{equation}
and ${\cal E}_{\mu \nu}$ is the projected 5D Weyl tensor. 
Here $\delta \rho_{{\cal E}}, \delta p_{{\cal E}}$ 
and $\delta q_{{\cal E}}$ are perturbations of ''Weyl fluid'';
\begin{equation}
-\delta {\cal E}^{\mu}_{\nu} = \kappa_4^2
\left(
\begin{array}{cc}
 -\delta \rho_{{\cal E}} & \delta q_{{\cal E},i} \\
 a_o^{-2} \delta q_{{\cal E},i}  & \delta p_{{\cal E}} \: \delta_{ij} 
 + \delta \pi_{{\cal E}\: ,ij}\\
\end{array}
\right).
\end{equation}

Substituting the solution for $\Omega$ (\ref{omega}), 
The metric perturbations and Weyl tensor are obtained as 
\begin{equation}
{\cal R}= 
-\int dm S(m) \left[ \frac{m}{\mu a_o} Z_1 \left(\frac{m}{\mu a_o}\right)
+ \frac{1}{3} \left( \frac{k}{\mu a_o} \right)^2 Z_0 \left(\frac{m}{\mu a_o}\right)
\right] e^{-i \omega T},
\end{equation}
\begin{eqnarray}
\kappa_4^2 \delta \rho_{{\cal E}}
&=& -\frac{2 k^4}{3 a_o^4} \int dm S(m) \mu^{-2} Z_0 \left(\frac{m}{\mu a_o}\right)
e^{-i \omega T},\nonumber\\
\kappa_4^2 \delta q_{{\cal E}}
&=& 
\frac{2 k^2}{3 a_o^3} \int dm S(m) 
\left[i \omega {\cal H} \mu^{-3} Z_0\left(\frac{m}{\mu a_o}\right) - H m \mu^{-3} Z_1 
\left(\frac{m}{\mu a_o}\right) \right], \nonumber\\
a_o^2 \kappa_4^2 \delta \pi_{{\cal E}}
&=& \int dm S(m) \frac{1}{3} (2k^2+3m^2) \mu^{-2} a_o^{-2} 
Z_0 \left(\frac{m}{\mu a_o}\right) e^{-i \omega T} \nonumber\\
&-& H^2 \mu^{-2} \int dm S(m) 
\left[ \frac{m}{\mu a_o} Z_1 \left(\frac{m}{\mu a_o}\right) 
-(k^2+2 m^2) \mu^{-2} a_o^{-2}Z_0\left(\frac{m}{\mu a_o}\right) \right]
e^{-i \omega T}
\nonumber\\
&+&
2 {\cal H}H \mu^{-2} \int dm S(m) i \omega m \mu^{-2} a_o^{-2} 
Z_1\left(\frac{m}{\mu a_o}\right) e^{-i \omega T}.
\end{eqnarray}
It should be noted that these equations have been derived already
in Ref.\cite{KS2}.
In Ref.\cite{KS2}, we solved the perturbations in Poincar\'e coordinate using the 
Randall-Sundrum gauge (see Appendix B). Of course the final result 
completely agrees. 

Now it is possible to write the equation $\delta p -c_s^2 \delta \rho=0$
in terms of the soltuion for $\Omega$. 
At low energies $H/\mu \ll 1$, the equaiton is simplified very much
because $-\kappa^2 {\cal H} = \kappa_4^2$ and we can neglect the 
terms proportional to $\dot{H}/{\cal H}^2$. 
For cosmological problems, it is natural to assume 
that the physical 3D wavelength of perturbations 
is much larger than the AdS curvature length at low energies. 
Thus we can neglect the terms suppressed by $k/\mu a_o$. 
Then we end up the 4D Einstein equation 
except for the equation
\begin{equation}
A+{\cal R} =-a_o^2 \kappa_4^2 \delta \pi_{{\cal E}},
\label{aniso}
\end{equation}
where 
\begin{eqnarray}
{\cal R} = -\int dm S(m) \left( \frac{m}{\mu a_o} \right) Z_1 \left(   
\frac{m}{\mu a_o} \right) e^{- i \omega \eta}, \nonumber\\
a_o^2 \kappa_4^2 \delta \pi_{{\cal E}} 
= \int dm S(m) \left(\frac{m}{\mu a_o} \right)^2 Z_0 
\left( \frac{m}{\mu a_o} \right) e^{- i \omega \eta}.
\label{R}
\end{eqnarray}
Here we implicitly assume that there is 
no dark radiation perturbation (see the conclusions section 
for a discussion of the dark radiation perturbation).
At super-horizon scales, the conservation of 
curvature perturbation on the hypersurface 
of uniform energy density; 
\begin{equation}
\zeta = {\cal R} -\frac{H^2}{\dot{H}} \left(  
\frac{\dot{{\cal R}}}{H} -A \right),
\end{equation}
can be shown without using the equation that relates 
${\cal R}$ and $A$ \cite{large}. However, in order to 
predict the Cosmic Microwave 
Background (CMB) anisotropy we need the relation between ${\cal R}$ 
and $A$ because the large scale Sachs-Wolfe effect is given by
\begin{equation}
\frac{\Delta T}{T}=\zeta+A-{\cal R}. 
\end{equation}
Thus we need to evaluate a Weyl anisotropic stress 
$\delta \pi_{{\cal E}}$ in order to address the CMB anisotropy. 

As for the tensor perturbations, if we assume 
\begin{equation}
\epsilon=\left( \frac{m}{\mu a_o} \right) \ll 1,
\end{equation}
and using the asymptotic formula for Hankel function
\begin{equation}
H_0^{(2)}(z) \sim -i \frac{2}{\pi} \left(\gamma+\log \frac{z}{2}\right)
+1, 
\quad \gamma: \mbox{Euler number},
\end{equation}
we get ${\cal R}=-A$ \cite{KS2}. Then we recover the 4D Einstein gravity. 
However, as in the case for the tensor
perturbation we should determine $S(m)$ to justify this 
assumption. The problem is the same as the tensor perturbation one. 
We used the same boundary and initial conditions. 
The cut-off of the KK mass is introduced in the numerical
calculations. As for the tensor perturbations, the obtained 
soltuion for $S(m)$ is localized well below the cut-off. 
In addition, for scalar perturbations, 
we neglect the terms suppressed  by $H/\mu$ and $k/\mu a_o$ 
and use the 4D Friedmann equation in performing the 
numerical calculations. 
Figure 4 shows the solution for $S(m)$. From these solutions 
we can construct the solution for metric perturbations. 
We recover the 4D behavior of the perturbations as expected 
(Figure 5). 
In particular, the metric perturbations satisfy the relation ${\cal R}=-A$. 
The behavior of metric perturbations does not depend on the solutions 
for $S(m)$; thus the recovery of 4D cosmology does not depend 
on the initial conditions.

\begin{figure}[h]
\begin{center}
\includegraphics[width=14cm]{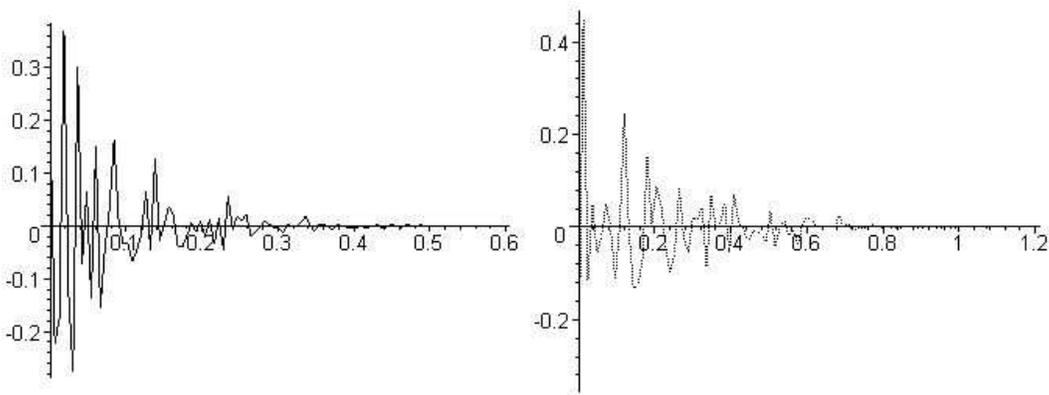}
\end{center}
\caption{We show two of the solutions for $S(m)$. 
The horizontal axis is $m/\mu$. The perturbations 
enter the horizon when $H/\mu=0.0001$. 
The universe is assumed to be radiation dominated. 
The initial condition is taken such that metric perturbation is 
constant at super horizon scales. The cut-off is introduced at
$m/\mu=1.2$.}
\end{figure}

\begin{figure}[h]
\begin{center}
\includegraphics[width=10cm]{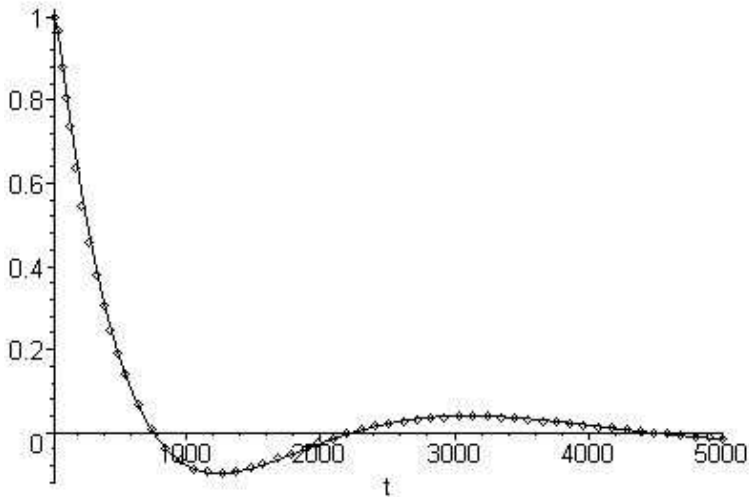}
\end{center}
\caption{The evolution of the metric perturbation ${\cal R}$ and 
$-A$ on the brane. 
The horizontal axis is the cosmic time.
The solid line is the prediction of 4D cosmology
with the same initial condition and the points represent the solutions
obtained by summing the Poincar\'e mode functions 
with the weight $S(m)$. The amplitude is normalized so that 
${\cal R}=1$ at the initial time.} 
\end{figure}

\section{The GN coordinate view}
In this section, we reconsider the evolution of scalar
perturbation in the GN coordinate.
As mentioned in section 3, the GN coordinate has 
a crucial disadvantage because it cannot cover the whole bulk 
spacetime due to the coordinate singularity. 
However it is still a good coordinate near the brane and 
we can understand what is happening near the brane in 
a very simple way. Thus it would be instructive to compare the 
approach used in this paper with the analysis in the 
GN coordinate. 

Let us consider the low energy/near brane approximation 
\cite{p27} \cite{p28}. 
The metric is simply given by 
\begin{equation}
ds^2=dy^2 + e^{-2 \mu y} (-dt^2 + a_o^{2} \delta_{ij} 
dx^i dx^j ).
\end{equation}
Note that this metric is valid only near the brane
\begin{equation}
e^{2 \mu y} < (\kappa^2 \rho/ \mu)^{-1}.
\label{renge}
\end{equation}
At late times, the derivative with 
respect to time is expected to be much weaker than the derivative 
with respect to $y$:
\begin{equation}
\ddot{\Omega} \ll \Omega'' \sim \mu^{-2} \Omega.
\end{equation}
With this approximation the master equation becomes 
\begin{equation}
\Omega''+2 \mu \Omega' +\mu^2 \Omega =0.
\end{equation}
Then the solution is given by
\begin{equation}
 \Omega= \Omega_0(x) \mu y e^{- \mu y} +C(x) e^{- \mu y}.
\label{solution}
\end{equation}
Here we give a weak dependence on brane coordinates 
to the constants of integration $\Omega_0(x)$ and 
$C(x)$. These dependences can be determined by imposing the boundary 
conditions. The junction condition is greatly simplified 
due to the fact $a'/a=n'/n$ at late times $\kappa^2 \rho/\mu \ll 1$. 
The junction conditions of equations (\ref{bomega}) are written as 
\begin{eqnarray}
a_o \kappa^2 \delta \rho & =& 
-\frac{k^2}{a_o^2} {\cal F} -3 H \dot{{\cal F}}, \nonumber\\
a_o \kappa^2 \delta p & = & \ddot{{\cal F}}+
2 H \dot{{\cal F}} +  \dot{H} {\cal F},
\end{eqnarray}
where we defined
\begin{equation}
{\cal F}= \Omega' + \mu \Omega.
\end{equation}
On imposing the equation of state $\delta p=c_s^2 \delta \rho$,
the equation for ${\cal F}$ is obtained as 
\begin{equation}
\ddot{{\cal F}} + (2 + 3 c_s^2) H \dot{{\cal F}} + 
\dot{H} {\cal F} +c_s^2 k^2 a_o^{-2} {\cal F}=0. 
\end{equation}
A point here is that ${\cal F}$ is written only in terms of 
$\Omega_0(x)$,
\begin{equation}
{\cal F}= \mu \Omega_0(x)
\end{equation}
Thus we cannot determine the function $C(x)$. It is a
natural consequence. We are basically solving the second order 
differential equation with respect to $y$. The boundary 
condition at the brane is insufficient for 
determining the solution. We should specify the boundary
condition in the bulk to determine the solution completely. 
However, the GN coordinate is not suitable for this 
purpose due to the existence of the coordinate singularity. 
Moreover, the gradient expansion method for solving the perturbation
cannot work for large $y$ (see equation (\ref{renge})). However, if we calculate 
the behavior of metric perturbations, the leading order term 
that comes from $C$ disappears; 
\begin{eqnarray}
A &=& \frac{1}{2 a_o} \mu^2 \Omega_0(x) +{\cal O} (\ddot{C}),\nonumber\\
{\cal R} &=& -\frac{1}{2 a_o}\mu^2 \Omega_0(x) +{\cal O}(\ddot{C}).
\label{metric}
\end{eqnarray}
Thus if the condition
\begin{equation}
{\cal O} (\ddot{C}) \ll \mu^2 \Omega_0
\label{conC}
\end{equation}
is satisfied, we can ignore the contribution from $C(x)$. 
Then the metric perturbations are solely given by $\Omega_0$. 
Hence the equation for ${\cal F}$ gives the 
evolution equation for metric perturbations
\begin{eqnarray}
A &=& -{\cal R}, \nonumber\\
\ddot{{\cal R}} &+& (4 +3 c_s^2) H \dot{{\cal R}} + 
(2 \dot{H}+3 H^2 +3 c_s^2 H^2) {\cal R}
+c_s^2 k^2 a_o^{-2} {\cal R}=0.
\label{4D}
\end{eqnarray}
These are nothing but the evolution equation in  
standard 4D cosmology. 

The crucial disadvantage of the GN coordinate is that 
there is no way to find the solution for $C$ as mentioned above. 
Hence it is impossible to justify the condition ({\ref{conC}}) 
that is needed to show the recovery of the 4D evolution
equations. And also the function $C$ could modify the relation 
between $A$ and ${\cal R}$
\begin{equation}
A+{\cal R}= {\cal O}(\ddot{C}).
\label{GNaniso}
\end{equation}
This equation is essential for calculating the CMB anisotropy. 
Thus we can say nothing about CMB anisotropy. 

Let us compare the above arguments with the analysis 
using mode functions in the Poincar\'e coordinate. 
At late times, ${\cal F}$ can be calculated as
\begin{equation}
{\cal F}=\mu \Omega_0= 2 \mu^{-2} \int dm S(m) m Z_1 \left(
\frac{m}{\mu a_o} \right)e^{-i \omega \eta} .
\end{equation}
From Eq.(\ref{R}), we can relate $\Omega_0$ to ${\cal R}$ as 
\begin{equation}
\Omega_0 = -2 a_o \mu^{-2} {\cal R},
\end{equation}
which agrees with Eq.(\ref{metric}).  
The difficulty that arises in the GN coordinate was that 
there is no way to determine $C$. In the numerical calculation done 
in Poincar\'e coordinate we did impose the boundary condition 
and initial conditions in the bulk. Hence the function $C$ 
was determined. 
The solution Eq.(\ref{solution}) indicates that $C$ is the value 
of $\Omega$ on the brane $\Omega_b=C$ 
(note that brane is located at $y=0$).
Thus $C$ is given in terms of the solution in Poincar\'e coordinate as 
\begin{equation}
\Omega_b=C=-2 \mu^{-2} a_o \int dm S(m) Z_0 \left(
\frac{m}{\mu a_o} \right) e^{-i \omega \eta}.
\end{equation}
We can also derive the equation that relates $A$ and ${\cal R}$ 
from Eqs.(\ref{aniso}) and (\ref{R});
\begin{eqnarray}
A+{\cal R} =-a_o^2 \kappa_4^2 \delta \pi_{{\cal E}}, \nonumber\\
a_o^2 \kappa_4^2 \delta \pi_{{\cal E}} 
= \int dm S(m) \left(\frac{m}{\mu a_o} \right)^2 Z_0 
\left( \frac{m}{\mu a_o} \right) e^{-i \omega \eta},
\end{eqnarray}
which should be compared with Eq.(\ref{GNaniso}).
The leading order behavior of Weyl anisotropic stress 
in $m/\mu a_o < 1$ contains the non-local term 
\begin{equation}
a_o^2 \kappa_4^2 \delta \pi_{{\cal E}}  
\propto \int dm S(m)  \left(\frac{m}{\mu a_o} \right)^2 
\log \left(\frac{m}{\mu a_o} \right) e^{- i \omega \eta}.
\label{log2}
\end{equation} 
Thus the correction $a_o^2 \kappa_4^2 \delta \pi_{{\cal E}}$ 
describes the 5D corrections that are caused by the propagation of 
perturbations into the bulk. This explains why the GN coordinate 
approach can tell nothing about this correction. 
We should determine the solution for the bulk perturbations 
completely to address the 5D correction. And also 
it is now understood that the physical meanings of 
the condition (\ref{conC}) is that the 5D effect that comes from
5D Weyl tensor $a_o^2 \kappa_4^2 \delta \pi_{{\cal E}}$ is 
negligible when determining the solution for metric perturbations
${\cal R}$ and $A$.

A point here is that we can quantitatively check the condition 
(\ref{conC}) using the numerical results done in the Poincar\'e coordinate.
Figure 6 shows the behavior of Weyl anisotropic stress 
from numerical solutions. The amplitude of the Weyl anisotropic stress
should be compared with the amplitude of ${\cal R}$ in Figure 5. 
The smallness of the amplitude of $a_o^2 \kappa_4^2 \delta \pi_{{\cal E}}$
compared with the amplitude of ${\cal R}$ indicates that 
the 5D effect is negligible when determining the solution for 
metric perturbations. This also quantitatively justify the condition
(\ref{conC}) in terms of the solutions in the Poincar\'e coordinate. 
Thus we can show the recovery of the 4D physics at low energies 
$H/\mu \ll 1$.

\begin{figure}[h]
\begin{center}
\includegraphics[width=10cm]{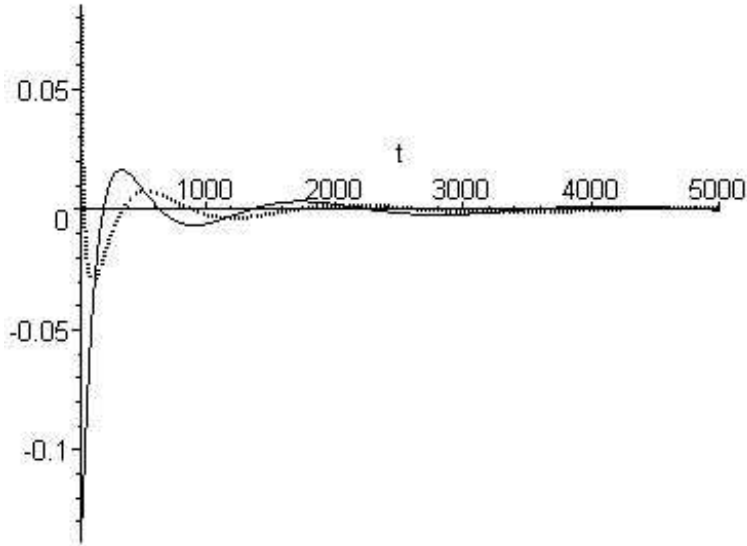}
\end{center}
\caption{Solutions for Weyl anisotropic stress $\kappa_4^2 a_o^2
\delta \pi_{{\cal E}}$. The amplitude is normalized so that
${\cal R}=1$ at the initial time (see Figure 5).
The horizontal axis is cosmic time. The solid line is the 
solution obtained from the solution for $S(m)$ in the 
left panel in Figure 4 and the dot line is that 
from in the right panel}
\end{figure}

\section{Conclusion}
We solved for the evolution of cosmological perturbations in a 
single-RS-brane world model. Assuming that the 
bulk is AdS spacetime, the general solutions in the Poincar\'e
coordinate were used to construct the solution. 
This allows us to find the solution throughout the infinite 
bulk spacetime. The junction conditions at the moving brane are 
imposed numerically. Then we obtained the solution for 
perturbations by solving the 5D bulk perturbations. 
We considered a late time evolution of perturbations. 
At late times $H/\mu \ll 1$, the behavior of the perturbations
agrees well with that in conventional 4D cosmology. 

Our result indicates that it is very difficult to 
find brane world corrections in CMB anisotropy 
in a single-RS-brane model. 
The AdS curvature scale is restricted to $\mu^{-1} < 
0.1$mm from table-top experiments of Newton's force
law. At high energies $H/\mu >1$, the perturbations that are 
relevant to the CMB physics stay at superhorizon scales. At superhorizon
scales, the curvature perturbation on hypersurface of 
uniform energy density is conserved even in a brane 
world model (note that there could be a correction from dark 
radiation perturbation; see below). Thus the curvature 
perturbation evolves in the same way as 4D cosmology. 
However, in order to predict the CMB anisotropy we need to 
know the anisotropic stress induced by 5D Weyl tensor in 
addition to the curvature perturbation. 
This could give a 5D effect to the CMB anisotropy.
Unfortunately, at the decoupling time when the 
CMB spectrum is determined, $H/\mu \ll 1$ is satisfied 
to extremely high accuracy. Our result indicates that 
this 5D effect is too small to be observed. 

However, there are still several possibilities to observe 
the brane world corrections. The first possibility is provided 
by the high energy corrections that arise when $H/\mu$ becomes large. 
This correction is 
particularly important for tensor perturbations because there is 
a possibility of directly proving the evolution of tensor perturbations
at high energies if we can observe the stochastic 
background of gravitational waves from inflation. 
A difficulty of the calculations at high energies is that
the behavior of the perturbations depends on the initial 
conditions. Thus we need an extra step to 
impose the initial conditions. A detailed numerical 
study of the high energy correction will be presented in a separate
publication \cite{HKT}. We should note that there has been a very interesting 
attempt to understand the high energy corrections in terms of 
AdS/CFT correspondence \cite{tanaka}. The logarithmic corrections 
$\log (m/\mu a_o)$ in Eqs.(\ref{log1}) and (\ref{log2}) can be understood
as corrections due to a coupling to the CFT. Although this approach can 
treat only mild corrections $H/\mu <1$, it is very important to 
understand the corrections analytically. It would be interesting to 
extend the analysis of Ref.\cite{tanaka} to scalar perturbations. 
 
Another possibility arises if we allow the 
existence of a black hole (BH) in the bulk. In the background 
spacetime, a bulk BH induces the so-called dark radiation
that modifies the evolution of the universe. If we consider 
the perturbation, dark radiation also has a perturbation and 
modifies the evolution of perturbations. An interesting point is
that this effect could be large even at low energies and 
it could give very interesting features in CMB anisotropy \cite{koyama}. 
Unfortunately, this dark radiation perturbation would be a 
non-normalizable mode in AdS spacetime \cite{YK}, \cite{KLMW}. 
However, there is a subtlety in the argument of the 
normalizability. It has been shown that the dark radiation 
perturbation corresponds to putting a small BH in the bulk \cite{YK}. 
We cannot treat the effect of the BH perturbatively 
near the event horizon even if the BH mass is small.  
So it is difficult to discuss the dark radiation perturbation 
in AdS spacetime background. We should carefully investigate 
this mode in AdS-Schwarzshild spacetime. 
Recently a master variable has been found in AdS-Schwarzshild 
spacetime \cite{IK}. Thus it would be very interesting to extend 
our analysis to AdS-Schwarzshild spacetime and investigate 
the late time behavior of perturbations. 

Finally, we should specify the initial conditions from inflation. 
For scalar perturbations, analysis has been done by assuming that the 
contribution from bulk perturbations can be neglected \cite{inf}. 
However we should carefully examine the validity of this 
assumption. Fortunately, during inflation, the equations are 
simplified significantly. Thus it is possible to analyze the behavior 
of perturbations analytically to some extent \cite{KLMW}. We hope that
the primordial fluctuations will also have some brane world 
signatures.

Quantitative analysis of 5D effects on cosmological perturbations in a 
single-RS-brane world has just begun and much things remain to be done. 
We hope that our study provides the first step toward detailed 
predictions of cosmological observations in this model. 

\section*{Acknowledgement}
I would like to thank Takashi Hiramatsu and Atsushi Taruya 
for discussions about numerical calculations and 
Roy Maartens and David Wands for discussions.

\appendix
\section{Junction condition for $\Omega$}
In the brane world, we should perturb the location of the 
brane as well as the perturbation in the bulk \cite{bend}. 
The 5D longitudinal gauge in GN coordinate completely fixes 
the gauge, thus there is no guarantee that the perturbed brane
is located at $y=0$. Thus we should go to the gauge where the 
perturbed brane is fixed at $y=0$. We call this coordinate 
the brane-GN coordinate. Under a scalar gauge transformation
\begin{equation}
t \rightarrow \bar{t} = t+\delta t, \quad 
y \rightarrow \bar{y} = y+ \delta y,\quad 
x \rightarrow \bar{x} = x+\delta x,
\end{equation}
metric perturbations transform as
\begin{eqnarray}
\bar{A}_{yy} &=& A_{yy}-\delta y',\quad 
\bar{A}_{y} = A_{y} +n \delta t' -\frac{1}{n} \dot{\delta y}, \quad
\bar{B}_{y} =  -\delta x' - \frac{1}{a^2} \delta y,\nonumber\\
\bar{A} &=& A-\dot{\delta t}-\frac{n'}{n} \delta y-\frac{\dot{n}}{n} 
\delta t,\quad
\bar{{\cal R}} = {\cal R}-\frac{a'}{a} \delta y -\frac{\dot{a}}{a} \delta 
t,\nonumber\\
\bar{E} &=& -\delta x,\quad
\bar{B} =  \frac{n^2}{a^2} \delta t -\dot{\delta x},
\end{eqnarray}
where
\begin{equation}
\label{pertmetric} g_{AB}= \left(
\begin{array}{ccc}
-n^2(1+2 A) & a^2B_{,i} & n A_y \nonumber\\
a^2 B_{,j} & a^2\left[ (1+2 {\cal R})\delta_{ij} + 2 E_{,ij} \right] &
a^2 B_{y,i} \nonumber\\
n A_y & a^2 B_{y,i} & 1+2 A_{yy} \nonumber
\end{array}
\right) \, .
\end{equation}
The conditions for the GN coordinate $\bar{A}_{yy}=\bar{A}_{y}
=\bar{B}_{y}=0$ give
\begin{eqnarray}
\delta y' &=& A_{yy}, \nonumber\\
\delta t' &=& -\frac{1}{n} A_y + \frac{1}{n^2} \dot{\delta y},\nonumber\\
\delta x' &=& -\frac{1}{a^2} \delta y.
\end{eqnarray}
The junction conditions in brane-GN coordinate are given by
\begin{eqnarray}
\bar{A}_b' &=& \frac{\kappa^2}{6} (2 \delta \rho +3 \delta p), \nonumber\\
\bar{R}_b' &=& -\frac{\kappa^2}{6} \delta \rho,\nonumber\\
\bar{B}_b' &=& \kappa^2 \frac{n^2}{a^2} \delta q,\nonumber\\
\bar{E}_b' &=& 0.
\end{eqnarray}
Using the junction condition for $\bar{E}$, we get
\begin{equation}
\bar{E}_b'=-\delta x_b'= \frac{1}{a_o^2} \delta y_b=0.
\end{equation}
Thus the brane location is not perturbed; that is $\delta y_b=0$. 
The metric perturbations in 5D longitudinal gauge can be 
regarded as the induced metric perturbations on the brane. 
There is a residual gauge freedom in $\delta t$ on the brane. Using 
this gauge freedom we can set 
\begin{equation}
\delta t_b=0.
\end{equation}
Then it is possible to write $\bar{A}_b', \bar{R}_b'$ and $\bar{B}_b'$ 
in terms of the metric perturbations in the 5D longitudinal 
gauge. Substituting the solutions for metric perturbations
in terms of a master variable, we can express the matter 
perturbations in terms of a master variable. 

\section{Scalar perturbation in Randall-Sundrum gauge}
An alternative way to solve scalar perturbations is to use 
the Randall-Sundrum gauge in the Poincar\'e coordinate. This was done in 
Ref.\cite{KS1}, \cite{KS2}.
We can start with the perturbed AdS spacetime in Poincar\'e coordinate; 
\begin{eqnarray}
ds^2= \left(\frac{1}{\mu z} \right)^2
\left(dz^2 - (1+2 \phi) d \tau^2 
+2 b_{,i} dx^i d \tau
+ \left((1 - 2 \hat{\Psi}) \delta_{ij}+ 
2 \hat{E}_{,ij} \right)dx^i dx^j \right), 
\nonumber\\
\end{eqnarray}
where $\phi,b,\hat{\Psi}$ and $\hat{E}$ are given by
\begin{eqnarray}
h = (\mu z)^2 \int \frac{d^3 k}{(2 \pi)^3}
\int d m \:\:
h (m)  Z_2 (m z) e^{-i \omega \tau}e^{i k x}, 
(h= \phi,b,\hat{\Psi},\hat{E}).
\label{B-2}
\end{eqnarray}
Here we used the transverse traceless gauge conditions
\begin{eqnarray}
&& \phi-3 \hat{\Psi} + \nabla^2 \hat{E} = 0, \nonumber\\
&& 2 \frac{d \phi}{d \tau} + \nabla^2 b =0, \nonumber\\
&& \frac{d b}{d \tau}+ 2 \hat{\Psi} -2 \nabla^2\hat{E}=0. 
\end{eqnarray}
Thus the coefficients $h (m)$ satisfy 
\begin{eqnarray}
\phi (m) &=& \frac{2 k^4}{3 m^2} \mu^{-2}  E(m), \nonumber\\
b (m) 
&=& -4 i \frac{\sqrt{k^2+m^2} \:\: k^2 \mu^{-2}}{3 m^2} 
E (m), \nonumber\\
\hat{\Psi}(m) &=& 
-\frac{k^2 \mu^{-2}}{3}  E (m), \nonumber\\
\hat{E} (m)&=& \frac{2 k^2 +3 m^2}{3 m^2} \mu^{-2} E(m),
\label{B-4}
\end{eqnarray}
where $E(m)$ is the arbitrary coefficient.
As in the case for 5D longitudinal gauge, it is possible 
to relate these solutions to matter perturbations on the
brane by imposing the junction conditions. The final result 
completely agrees with the one derived using
a master variable with the identification $E(m)=S(m)$.

\section{Numerical accuracy}
In this section we show the accuracy of the numerical 
calculations. We have checked the accuracy of the 
junction condition. For the tensor perturbation 
we evaluate equation (\ref{boun}) using the solution
for $h(m)$ (Figure C1). For scalar perturbation we 
evaluate $\kappa^2 \mu (\delta \rho -c_s^2 \delta p)$ 
using the solution for $S(m)$ (Figure C2). 
In the figures, we make the equations dimensionless 
using $\mu$. 

\begin{figure}[h]
\begin{center}
\includegraphics[width=11cm]{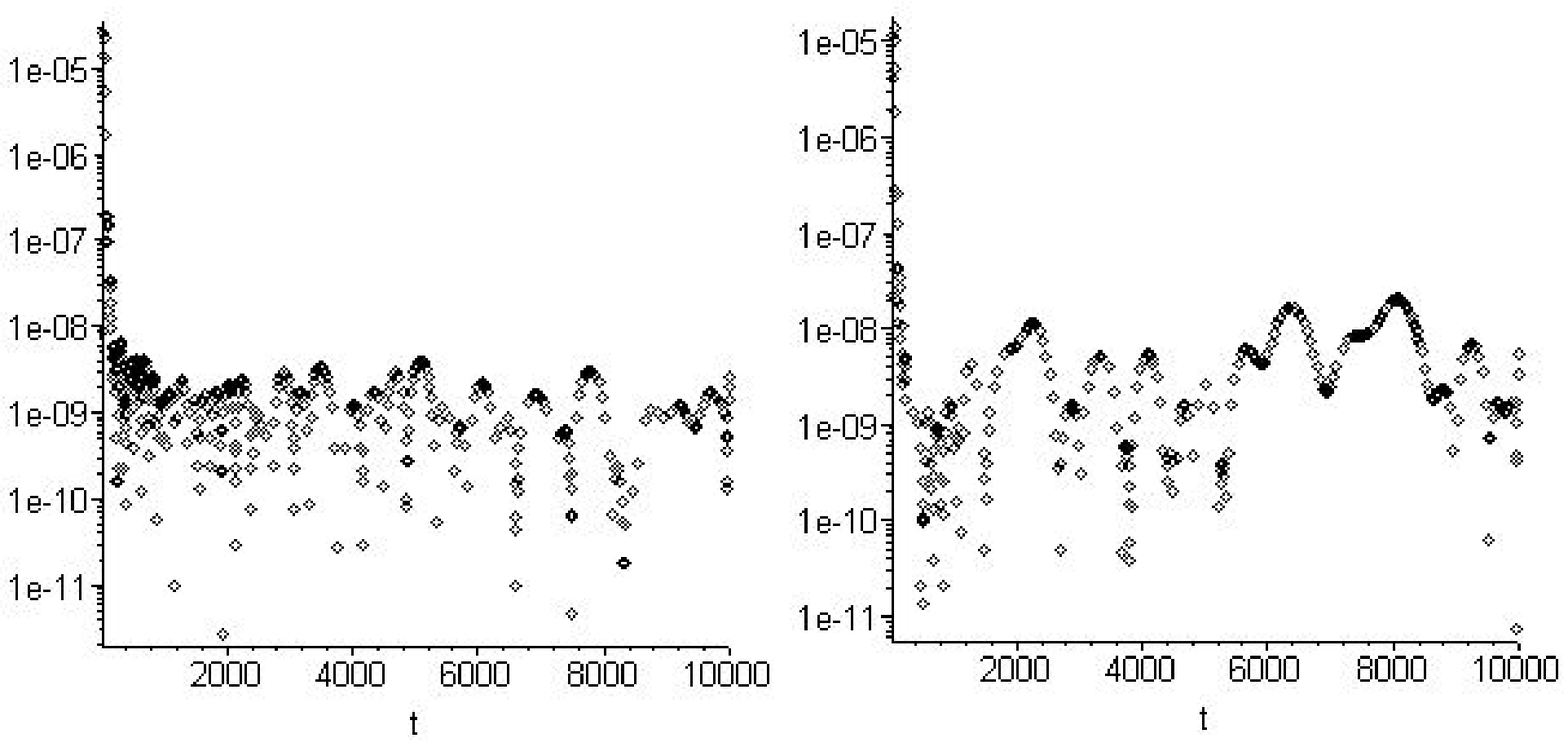}
\end{center}
\caption{The evaluation of Eq.(\ref{boun}) multiplied $\mu^{-1}$. 
The left (right) panel shows the result for $h(m)$ in the left 
(right) panel of Figure 1.}
\end{figure}

\begin{figure}[h]
\begin{center}
\includegraphics[width=11cm]{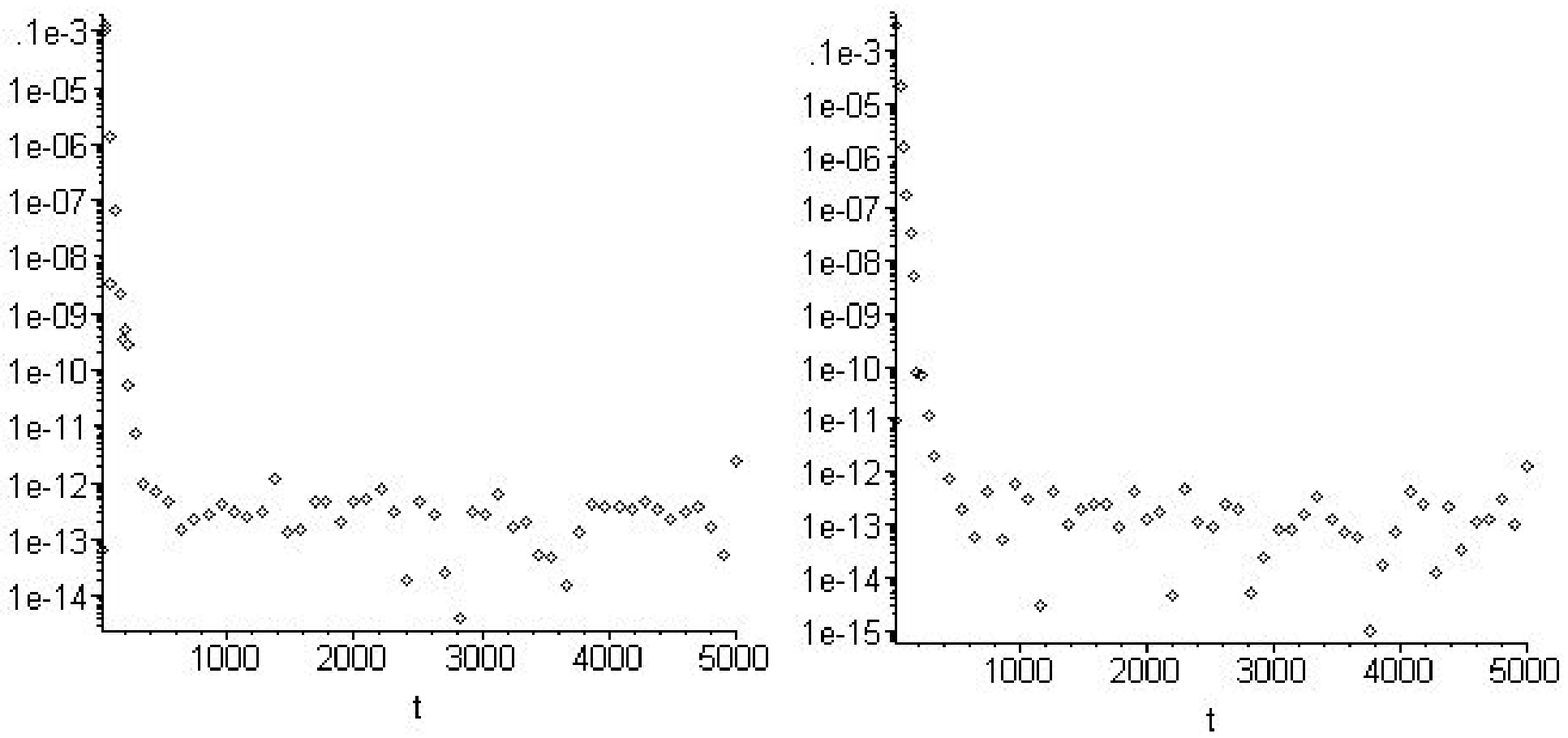}
\end{center}
\caption{The evaluation of $\kappa^2 \mu (\delta p-c_s^2 \delta \rho$)
multiplied by $\mu^{-2}$.  
The left (right) panel shows the result for $S(m)$ in the left 
(right) panel of Figure 4.}
\end{figure}

We need appropriate references 
to compare with the numerical violation of the junction 
condition. We define the "normalized" error of the junction
condition by dividing equation (\ref{boun}) by the representative
term in the junction condiiton. For the tensor perturbation
we divided the junction condition by the first term of 
equation (\ref{boun}) (Figure C3). For the scalar perturbation, 
we divided the equation $\kappa^2 \mu (\delta \rho -c_s^2 \delta p)$ 
by
\begin{equation}
\int dm S(m) \omega^2 \frac{m}{\mu a_o^3} Z_1 \left(  
\frac{m}{\mu a_o} \right) e^{-i \omega T}.
\end{equation}
which has the typical amplitude of the terms in the equation 
$\kappa^2 \mu (\delta \rho -c_s^2 \delta p)$. The result is shown in 
Figure C4. By construction, the 
normalized error is dimensionless and it measures the 
violation of the junction conditions. Note that
the appearance of peaks of large errors is caused by 
the fact that the denominator becomes $0$; thus it 
is an artifact of the definition of the normalized 
error.  

\begin{figure}[h]
\begin{center}
\includegraphics[width=11cm]{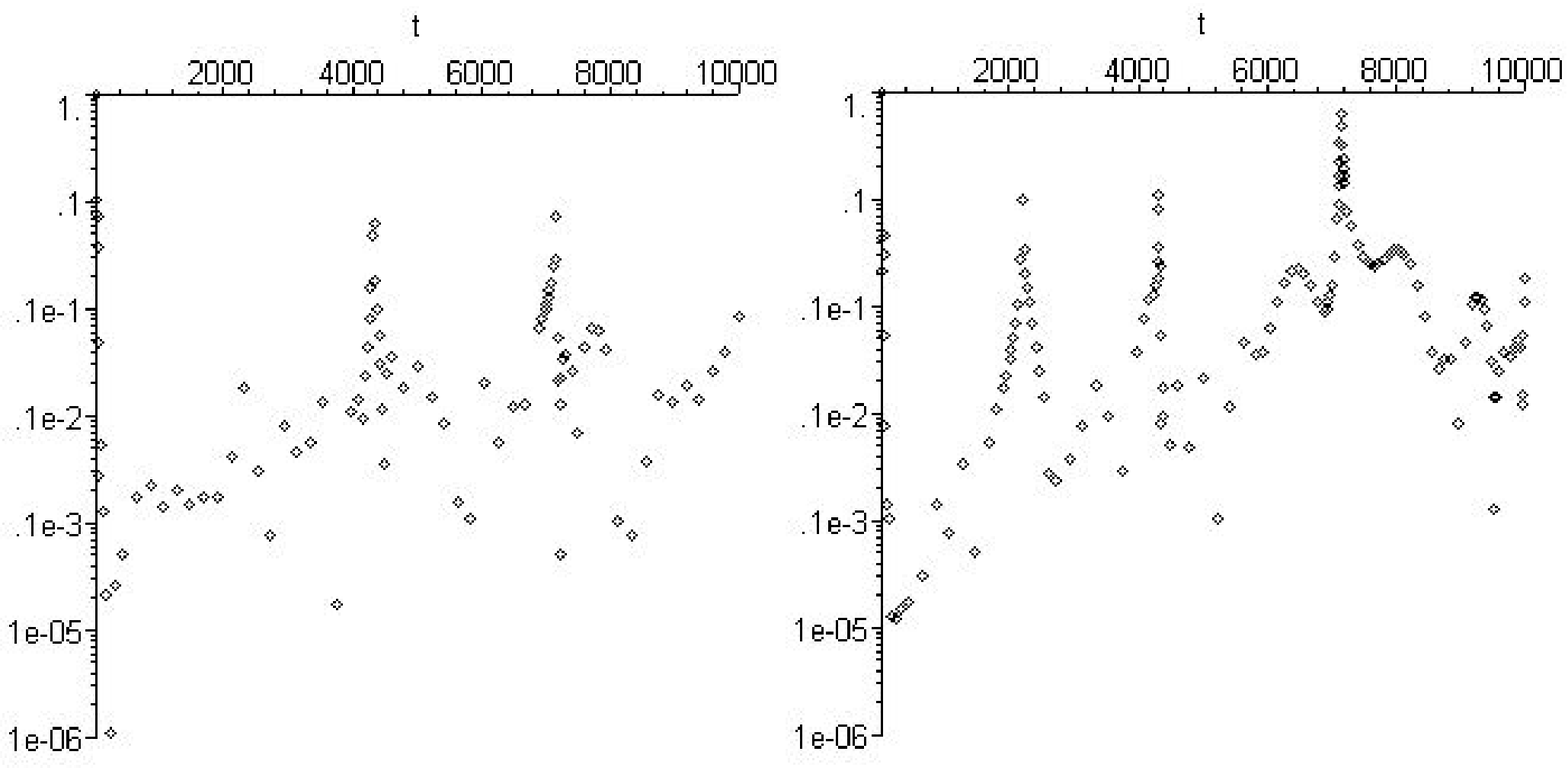}
\end{center}
\caption{The normalized error for the tensor perturbation. 
The left (right) panel shows the result for $h(m)$ in the left 
(right) panel of Figure 1.}
\end{figure}

\begin{figure}[h]
\begin{center}
\includegraphics[width=11cm]{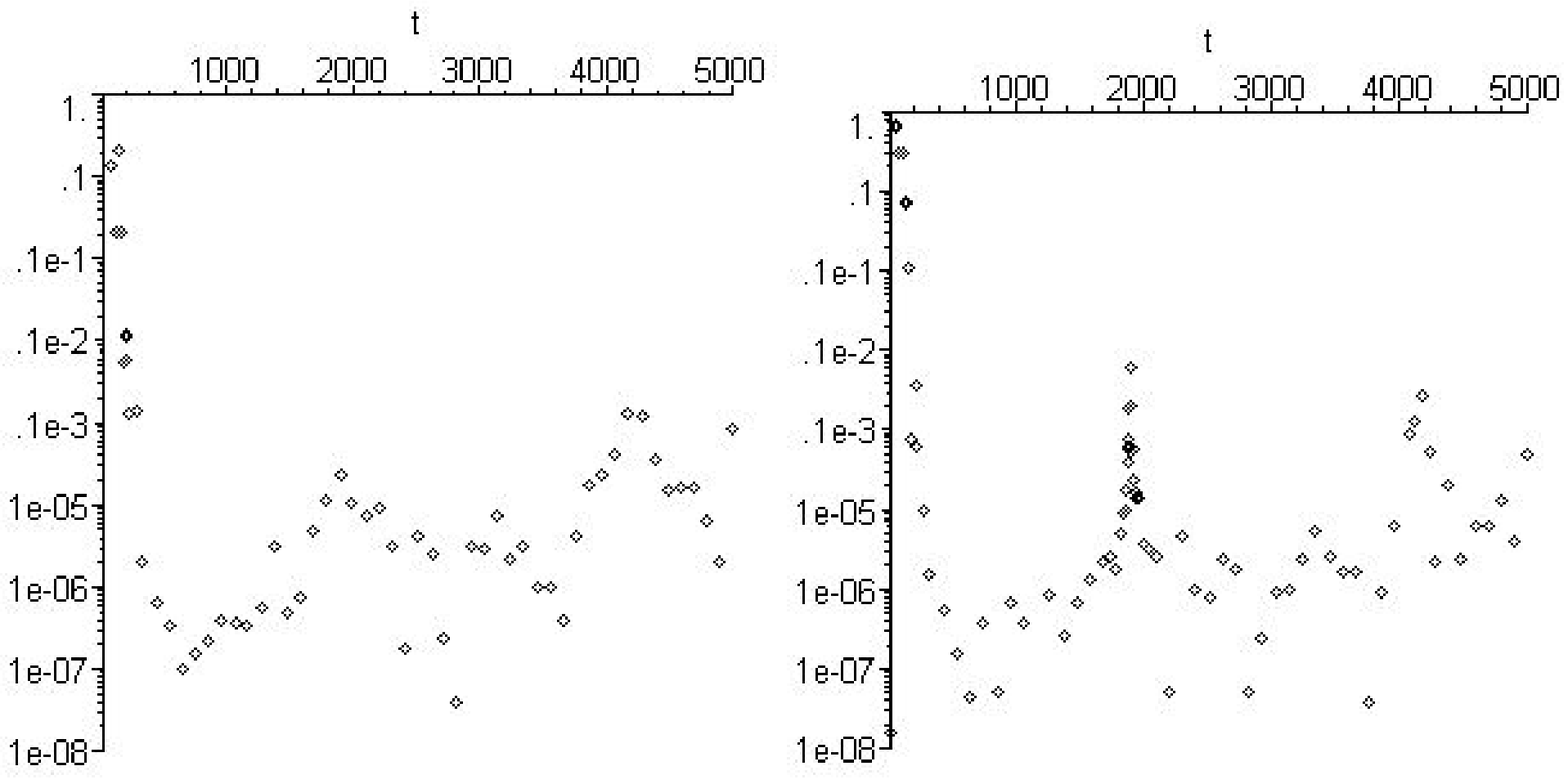}
\end{center}
\caption{The normalized error for the scalar perturbation. 
The left (right) panel shows the result for $S(m)$ in the left 
(right) panel of Figure 4.}
\end{figure}

For comparison, the deviation from the 4D solution with the 
same initial conditions is shown in Figure C5 for the tensor perturbation. 
Within the accuracy of the numerical calculations, the 
solution obtained cannot be distinguished from the 4D solution. 

\begin{figure}[h]
\begin{center}
\includegraphics[width=7cm]{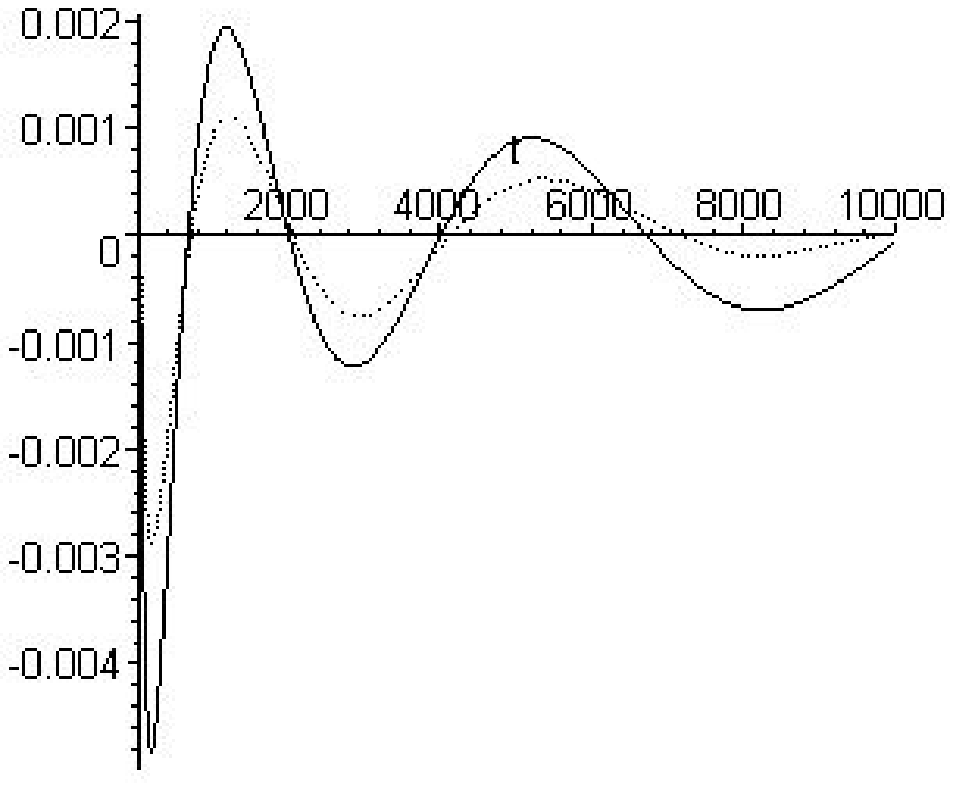}
\end{center}
\caption{The deviation from the 4D solution with the same initial 
condition. The solid line corresponds to the solution derived 
from $h(m)$ in the left panel of Figure 1 and the dotted line 
corresponds to that in the right panel.}
\end{figure}


\section*{References}
\begin{thebibliography}{99}

\bibitem{review}
Maartens R, {\it  Brane World Gravity}, 2003, Living Rev. Rel. [gr-qc/0312059];
Brax Ph, van de Bruck and Davis AC, 
{\it Brane World Cosmology}, 2003, Rept. Prog. Phys. [hep-th/0404011];
Langlois D, {\it Brane Cosmology: An Introduction},
2003, Prog. Theor. Phys. Suppl. {\bf 148}, 181 [hep-th/0209261]

\bibitem{RS} 
Randall L and Sundrum R, {\it An alternative to compactification}, 1999, Phys. Rev. 
Lett. {\bf 83} 4690 [hep-th/9906064], see also
Randall L and Sundrum R, {\it A large mass hierarchy from small extra dimension}, 
1999, Phys. Rev. Lett. {\bf 83} 3370 [hep-th/9905221]

\bibitem{B1} Bin\'etruy P, Deffayet C, Ellwanger U and Langlois D, {\it Brane 
cosmological evolution on a bulk with a cosmological
constant}, 2000, Phys. Lett. {\bf B477}, 285 
[hep-th/9910219]

\bibitem{Kraus} Kraus P, {\it Dynamics of anti-de Sitter 
domain walls}, 1999, JHEP, {\bf 12} 011 [hep-th/9910149]

\bibitem{B2} Ida D, {\it Brane world cosmology}, 
2000, JHEP, {\bf 09} 014 
[gr-qc/9912002]

\bibitem{B3} Mukohyama S, Shiromizu T, Maeda K, 
{\it Global structure of exact cosmological solutions in the brane world}, 2000, Phys. 
Rev. {\bf D62}, 024028 [hep-th/9912287]

\bibitem{Mu}Mukohyama S, {\it Gauge invariant gravitational perturbations of maximaly 
symmetric spacetime}, 2000, 
Phys. Rev. {\bf D62}, 084015
[hep-th/0004067]; {\it Perturbation of junction condition and doubly gauge invariant 
variables}, 2000, 
Class. Quant. Grav. {\bf 17}, 4777 [hep-th/0006146]; 


\bibitem{KIS} Kodama H, Ishibashi A and Seto O, {\it Brane-world cosmology, gauge 
invariant formalism for perturbation},2000, Phys. Rev. {\bf D62}, 064022
[hep-th/0004160]

\bibitem{P3} Maartens R, {\it Cosmological dynamics on the brane}, 2000, Phys. Rev. 
{\bf D62}, 084023 
[hep-th/0004166]

\bibitem{P4} Langlois D, {\it Brane cosmological perturbations}, 2000, Phys. Rev {\bf 
D62}, 126012
[hep-th/0005025]; {\it Evolution of cosmological
perturbations in a brane universe}, 2001, 
Phys. Rev. Lett. {\bf D86}, 2212 [hep-th/0010063]

\bibitem{BDBL} van de Bruck C, Dorca M, Brandenberger R and 
Lukas A, {\it Cosmological perturbations in brane world theories~: formalism}, 2000, 
Phys. Rev. {\bf D62},
123515 [hep-th/0005032] 

\bibitem{KS1} Koyama K and Soda J, {\it Evolution of cosmological perturbations in the 
brane world},
2000, Phys. Rev. {\bf D62}, 123502 [hep-th/0005239]

\bibitem{P7} Langlois D, Maartens R and Wands D. 
{\it Gravitational waves from inflation on the brane}, 
2000, Phys. Lett. {\bf B489}, 259 [hep-th/0006007]

\bibitem{p8} Gordon C and Maartens R, {\it 
Density perturbations in the brane-world},
2001, Phys. Rev. {\bf D63}, 044022 [hep-th/0009010]

\bibitem{p9} Bridgman H, Malik K and Wands D, {\it 
Cosmic vorticity on the brane}, 2001, Phys. Rev. {\bf D63}, 084012
[hep-th/0010133]

\bibitem{DDK} Deruelle N, Dolezel T and Katz J, {\it 
Perturbations of braneworlds}, 
2001, Phys. Rev. {\bf D63} 083513 [hep-th/0010215]

\bibitem{large} Langlois D, Maartens R, Sasaki M and Wands D, 
{\it Large-scale cosmological perturbations on the brane}, 
2001, Phys.Rev. {\bf D63}, 084009 [hep-th/0012044]

\bibitem{p12} Dorca M and van de Bruck C, {\it Cosmological perturbations in brane 
worlds~: brane bending and anisotropic stresses}, 2001, Nucl. Phys. {\bf B 605}, 215
[hep-th/0012116]

\bibitem{KH} Kodama H, {\it 
Behavior of cosmological perturbations in the brane world model}, 
2000,[hep-th/0012132] 

\bibitem{p14} Maartens R, {\it Geometry and dynamics of the brane-world}, 2001 
[gr-qc/0101059] 

\bibitem{p15} Mukohyama S, {\it Integro-differential equation for brane-world 
cosmological perturbations}, 2001, 
Phys. Rev. {\bf D64} 064006 [hep-th/0104185]

\bibitem{p16} Bridgman H, Malik K and Wands D, {\it Cosmological perturbations in the 
bulk and on the brane}, 
2002, Phys. Rev. {\bf D65} 043502 [astro-ph/0107245] 

\bibitem{KS2} Koyama K and Soda J, {\it Bulk gravitational field and cosmological 
perturbations on the brane}, 
2002, Phys. Rev. {\bf D65} 023514 [hep-th/0108003]

\bibitem{p18} Gorbunov D.S, Rubakov V.A and Sibiryakov S.M, {\it 
Gravity waves from inflating brane or mirrors moving in ads5}, 
2001, JHEP {\bf 0110} 015 [hep-th/0108017]

\bibitem{p19} Barrow J.D and Maartens R, {\it 
Kaluza-Klein anisotropy in the CMB}, 2002, 
Phys. Lett. {\bf B532}, 153 [gr-qc/0108073] 

\bibitem{p20} Leong B, Dunsby P, Challinor A and Lasenby A, 
{\it 1+3 covariant dynamics  of scalar perturbations in braneworlds}, 2002, Phys. Rev. 
{\bf D65} 104012 [gr-qc/0111033] 

\bibitem{p21} Deffayet C, {\it On brane world cosmological perturbations}, 2002, Phys. 
Rev {\bf D66} 103504 [hep-th/0205084] 

\bibitem{p22} Soda J and Koyama K, {\it Cosmological perturbations in brane world}, 
2002, [hep-th/0205208]

\bibitem{p23} Riazuelo A, Vernizzi F, Steer D and Durrer R, {\it 
Gauge invariant cosmological perturbation theory for braneworlds}, 2002 
[hep-th/0205220] 

\bibitem{p24} Leong B, Challinor A, Maartens R and Lasenby A, {\it Braneworld tensor 
anisotropies in the CMB}, 2002, 
Phys. Rev. {\bf D66} 104010 [astro-ph/0208015]

\bibitem{p25} 
Kobayashi T, Kudoh H and Tanaka T, {\it Primordial gravitational waves in inflationary 
braneworld}, 2003, Phys. Rev. {\bf D68} 044025  [gr-qc/0305006]. 

\bibitem{p26}
Hiramatsu T, Koyama K and Taruya A, {\it 
Evolution of gravitational waves from inflationary brane world:
numerical study of high energy effects}, 2004, 
Phys. Lett. {\bf B578} 269 [hep-th/0308072]

\bibitem{p27}
Easther R, Langlois D, Maartens R and Wands D, 
{\it Evolution of gravitational waves in Randall-Sundrum cosmology},
2003, JCAP {\bf 0310} 014 [hep-th/0308078]

\bibitem{p28}
Battye R, van de Bruck C and Mennim A, {\it
Cosmological tensor perturbations in the Randall-Sundrum model:
Evolution in the near brane limit}, 2004, 
Phys. Rev. {\bf D69} 064040 [hep-th/0308134]

\bibitem{p29} Ichiki K and Nakamura K, {\it 
Causal structure and gravitational waves in brane world cosmology}, 2003
[hep-th/0310282]; 
{\it Stochastic gravitational wave background in brane world cosmology}
,2004 [astro-ph/0406606]

\bibitem{tanaka} Tanaka T, {\it AdS/CFT correspondence in a Friedmann-Lemaitre-
Robertson-Walker brane}, 2004 [gr-qc/0402068]

\bibitem{p31}
Binetruy P, Bucher M and Carvalho, C, {\it
Models for the brane bulk interaction: toward understanding 
brane world cosmological perturbations}, 2004 [hep-th/0403154]

\bibitem{review-p}
Maartens R, {\it Brane-world cosmological perturbations}, 
2004 [astro-ph/0402485]; Deruelle N, {\it Linearized gravity
on branes: from newton's law to cosmological perturbations},
2003 [gr-qc/0301036]

\bibitem{koyama}
Koyama K, {\it Cosmic Microwave Radiation Anisotropy in brane worlds}, 2003, 
Phys. Rev. Lett {\bf 91}, 221301 [astro-ph/0303108]

\bibitem{CMB}
Rhodes C.S., van de Bruck C, Brax Ph and Davis A.C.,
{\it CMB anisotopies in the presence of extra dimensions}, 
2003, Phys. Rev {\bf D68}, 083511

\bibitem{HKT} Hiramatsu T, Koyama K and Taruya A,
in preparation

\bibitem{SMS} Shiromizu T, Maeda K and Sasaki M, {\it 
The Einstein equations on the 3-brane world},2000, 
Phys. Rev. {\bf D62} 024012 [gr-qc/9910076]

\bibitem{YK} Yoshiguchi H and Koyama K, 2004,
Phys. Rev. {\bf D70} 043513 [hep-th/0403097]

\bibitem{KLMW} Koyama K, Langlois D, Maartens R and 
Wands D, Preprint hep-th/0408222

\bibitem{IK} Kodama H and Ishibashi A, {\it A master equation
for gravitational perturbations of maximally symmetric 
black holes in higher dimensions}, 2003, 
Prog. Theor. Phys. {\bf 110}, 701 [hep-th/0305147]

\bibitem{inf} Maartens R, Wands D, Bassett B and Heard I, 
{\it Chaotic inflation on the brane},2000, 
Phys. Rev. {\bf D62}, 041301 [hep-th/9912464]


\bibitem{bend} Garriga J and Tanaka T, {\it 
Gravity in the Randall-Sundrum brane world}, 2000
Phys. Rev. Lett. {\bf 84}, 2778 [hep-th/9911055]
\end{thebibliography}
\end{document}